\documentclass[prd,twocolumn,showpacs,floatfix,amsmath,amssymb,floatfix]{revtex4}
\usepackage{graphicx,color,dcolumn,booktabs,bm}
\usepackage{longtable,lscape}
\usepackage{txfonts}
\usepackage{overpic}
\usepackage{amssymb}
\usepackage{indentfirst}
\usepackage{feynmf}   
\usepackage{slashed}  
\usepackage{cases}
\usepackage{color}
\usepackage{multirow}
\usepackage{graphicx,color,dcolumn,booktabs,bm}
\usepackage{epsfig,dsfont,amssymb,amsmath,amsfonts,amsbsy,mathrsfs}

\graphicspath{{Figures/}} %

\begin{document}

\title{ Study on the strong decays of $\phi(2170)$ and a grand expectation for the future charm-tau factory }

\vspace{1cm}

\author{ Hong-Wei Ke$^1$\footnote{khw020056@hotmail.com} and
        Xue-Qian Li$^2$\footnote{lixq@nankai.edu.cn}  }

\affiliation{  $^{1}$ School of Science, Tianjin University, Tianjin 300072, China \\
  $^{2}$ School of Physics, Nankai University, Tianjin 300071, China }

\vspace{12cm}

\begin{abstract}

The present data imply that $\phi(2170)$ may not be an excited state of $\phi$, but
is a four quark state with $ss\bar s \bar s$
constituents. Furthermore, there are no two mesons of $s\bar s$
available to form a molecule which fits the mass spectrum of
$\phi(2170)$, thus we suggest it should be an $ss\bar s \bar s$
tetraquark state. In this scenario, we estimate its decay rates
through the fall-apart mechanism. Our theoretical estimates indicate that
its main decay modes should be $\phi(2170)$ into $\phi f_0(980)$, $
h_1\eta$, $ h_1\eta'$, $K_1(1270)K$ and $K_1(1400)K$. Under this
hypothesis the modes $\phi(2170)\to K^*(890)^0\bar K^*(890)^0$,
$K^+K^-$ and $K^0_LK^0_S$ should be relatively suppressed. Since the width of
$h_1$ is rather large, at present it is hard to gain precise data on
$BR(\phi(2170)\to h_1\eta)$ and  $BR(\phi(2170)\to h_1\eta')$ whose measurements may be crucial
for drawing a definite conclusion about the inner assignment of $\phi(2170)$.
We lay our expectation to the proposed
charm-tau factory which will have much larger luminosity and better capacities.


\end{abstract}

\maketitle

\section{Introduction}

Very recently a meson  $\phi(2170)$ comes into the view of
researchers because it may be a special exotic state. It is
observed via its decay into $\phi+ f_0(980)$
\cite{Aubert:2006bu,Aubert:2007ur,Aubert:2007ym,Lees:2011zi,Ablikim:2007ab,Shen:2009zze,Ablikim:2014pfc,Ablikim:2017auj,Shen:2009mr}
meanwhile some possible final states $K^{*0}K^{\pm}\pi^{\pm}$ and
$K^{*0}\bar K^{*0}$ are not seen. If it were a normal meson i.e.
an excited state of $\phi$ the decay portals into $K^+K^-$ and
$K^0_LK^0_S$ would be preferred as $\phi$ does. Moreover, even the
channel $K^{*0}\bar K^{*0}$ should also be seen since a sufficient
phase space is available. Furthermore in Ref.\cite{Coito:2009na}
the theoretical evaluation on the total width obviously conflicts
to data if $\phi(2170)$ is a normal meson. A reasonable
interpretation is needed. It is suggested that the observed
$\phi(2170)$ could be a molecular state of $\Lambda \bar
\Lambda$\cite{Dong:2017rmg} or a tetraquark
state\cite{Ali:2011qi}. In reference\cite{Takeuchi:2017qyg} the
author thinks that $\phi(2170)$ is an excited $q\bar q s\bar s$
tetraquark ($q=u, d$ ). But this assignment is questionable
because no ground $q\bar q s\bar s$ tetraquark has ever been
observed.

Being hinted by the decay mode $\phi(2170)\to\phi f_0(980)$, a
natural conjecture is that $\phi(2170)$ may be a four quark state
with $ss\bar s \bar s$ constituents. There are two choices: a
molecular state or a tetraquark. However, there are no two mesons
with $s\bar s$ constituents available to form a molecular hadron
which fits the mass spectrum of $\phi(2170)$, thus we turn to
suggest that it is an  $ss\bar s \bar s$ tetraquark state. This
conjecture was also considered by the authors of
Ref.\cite{Chen:2018kuu,Wang:2006ri}.

At the end of last century a stimulating question was raised: did
multiquark states indeed exist in nature, because in the primary
paper Gell-Mann predicted them along with the simplest assignments
of $q\bar q$ for mesons and $qqq$ for
baryons\cite{GellMann:1964nj}. The first proposed pentaquark of
$qqqq\bar s$ with unusual $B=1$ and $S=1$ quantum numbers would
definitely be a multiquark state. In that assignment except $\bar
s$, all other quarks are light ones (u or d types). The passion of
detecting such pentaquarks was very high, however, after hard and
desperate search, such pentaquarks were never observed
experimentally. The despair discourages researchers who decided to
give up. But following conduction of more accurate experiments and
innovated skills of analysis, many exotic mesons have been
measured. They are proposed to be four-quark states
(molecular states or tetraquarks states)\cite{Abe:2007jn,
Choi:2005,Choi:2007wga,Aubert:2005rm,Ablikim:2013emm,
Ablikim:2013wzq,Ablikim:2013mio,Liu:2013dau,Collaboration:2011gja,Ablikim:2014dxl},
later two pentaquarks were observed by the LHCb
collaboration\cite{Aaij:2015tga}. It validates the suggestion
about existence of multi-quark states. However, we have observed
that all the discovered multi-quark states contain at least one
heavy quark ($c$ or $b$). This may hint that the existence of
heavy quarks in the multi-quark states is fatal \cite{Li:2014gra}.
Is that the conclusion of the story?
$\phi(2170)$ which comes into our attention recently,
could be identified as a four quark state with
$ss\bar s\bar s$ constituents. Even though it is true, the early
allegation might not be completely subverted because the mass of
$s$-quark resides between that of very light quark and the
supposed ``heavy" charm quark and the rest constituents in the
exotic state are all $s$-flavor ($\bar s$) with ``middle" mass.

A naive analysis may provide us a support about this conjecture.
The masses of $\Omega$ and $\phi$ which consist of three $s$
quarks and an $s\bar s$ respectively, are 1672 MeV and 1020 MeV.
It implies the s-quark mass to be around 500$\sim$600 MeV, thus a
simple estimate on the mass of  $ss\bar s\bar s$ tetraquark state
should fall in a region close to the mass of $\phi(2170)$. If the
assignment is true $\phi(2170)$ is indeed a tetraquark with a
single flavor of strangeness.

No doubt, it is absolutely important to get a better understanding
about the inner structure of $\phi(2170)$. Since it only possess
$s$ flavor, its decays would be dominated by the modes where the
final states mainly contain $s$ flavors. Let us turn to
investigate the mechanism which governs the strong decay of
$\phi(2170)$. It is the so called ``fall-apart"
mechanism\cite{Jaffe:1976ih,Kim:2017yvd}.

In Refs.\cite{Jaffe:1976ih,Kim:2017yvd}
they suggested that the fall-apart mechanism induces the main decay modes of the tetraquark state.
By this mechanism the constituents in a tetraquark
are rearranged into two color singular pairs by exchanging soft gluons and then simply
fall apart into two mesons.
In this work  we will employ this
mechanism to study the decays of $\phi(2170)$.

This paper is organized as follows: after the introduction, in
section II we will explore the decays of $\phi(2170)$. Section III
is devoted to our conclusion and discussions.

\section{Fall-apart decays of $\phi(2170)$}
\begin{figure}
\begin{center}
\scalebox{0.4}{\includegraphics{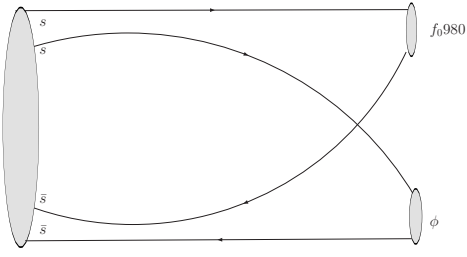}}
\end{center}
\caption{The Feynman diagram for $ss\bar s \bar s\to \phi
f_0(980)$ transitions via the fall-apart mechanism}\label{t1}
\end{figure}

1. Since  $\phi(2170)$ of $J^P=1^-$ is supposed to be an $ss\bar s\bar s$
tetraquark which is in a diquark-antiquark configuration, its spin state is
\begin{eqnarray}
|J,J_{12},J_{34}\rangle=|1,1,1\rangle,
\end{eqnarray}
where $J$ is the spin of the tetraquark $ss\bar s\bar s$, $J_{12}$
is the spin of $ss$ and  $J_{34}$ is the spin of $\bar s \bar s$.
The orbital angular momentum between diquark and anti-diquark is
1, i.e. in $p$-wave for guaranteeing the parity to be negative.
Moreover, since the $C$-parity of $\phi(2170)$ is odd so the spin
configuration of the tetraquark is fully determined as
\begin{eqnarray}
|1,1\rangle=&&\frac{1}{\sqrt{2}}(|1,1\rangle_{ss}|1,0\rangle_{\bar
s \bar s}-|1,0\rangle_{ss}|1,1\rangle_{\bar s\bar s}).
\end{eqnarray}

The color configuration is $|1,\bar 3,\bar 3\rangle$ which can be
written as\cite{Kim:2017yvd}
\begin{eqnarray}
\frac{1}{\sqrt{48}}\varepsilon_{abd}\varepsilon^{aef}(s^bs^d)(\bar
s_e\bar s_f).
\end{eqnarray}
Note, the spin configuration of the tetraquark $ss\bar s\bar s$ is in
the diquark and antidiquark spin bases. When it decays via the
fall-apart mechanism, one needs to switch a pair quark-anti-quark around and
rearrange their spins and colors to make proper
combinations for the two mesons in the final state.

2. Now let us study the decay of $\phi(2170)$ via the fall-apart mechanism. Apparently
the two-body finial states with $s$ wave is preferred if it is allowed.
Since the $J^{PC}$ of $\phi(2170)$ is $1^{--}$ tetraquark
$ss\bar s\bar s$ can fall apart into two mesons with the quantum
number assignments as
 $1^{--}$ and $0^{++}$ or
$1^{+-}$ and $0^{-+}$
\begin{eqnarray}
|1,J_z\rangle=&&\frac{1}{2}(|1,m\rangle_{13}|0,0\rangle_{24}+|0,0\rangle_{13}|1,m\rangle_{24}\nonumber
\\&&+|1,m\rangle_{14}|0,0\rangle_{23}+|0,0\rangle_{14}|1,m\rangle_{23}),
\end{eqnarray}
with $J_z=m$.

$\phi(2170)$ can also fall apart into two mesons with the quantum numbers $1^{++}$
and $1^{--}$
\begin{eqnarray}
|1,J_z\rangle=&&\frac{1}{\sqrt{2}}(\sum_{m_{13}m_{24}}C_{m_{13}m_{24}}|1,m_{13}\rangle|1,m_{24}\rangle
\nonumber
\\&&+\sum_{m_{14}m_{23}}C_{m_{14}m_{23}}|1,m_{14}\rangle|1,m_{23}\rangle),
\end{eqnarray}
with ${ {\mathbf{J}}}_{13}={ \mathbf{J}}_1+{\mathbf{J}}_3$, ${
\mathbf{J}}_{24}={\mathbf{J}}_2+{\mathbf{J}}_4$, $m_{13}$ and
$m_{24}$ are their projections along Z-axis, while
$J_z=m_{13}+m_{24}$. $C_{m_{13}m_{24}}$ and $C_{m_{14}m_{23}}$ are
corresponding C-G coefficients.

One also notices: the $I^G$ of $\phi(2170)$ is $0^-$, for such strong
OZI-allowed decays  the
two finial mesons should more favorably be in $I^G=0^-$ and
$0^+$ respectively, of course, the combination of $1^-,1^+$ could also
work, but naively may be suppressed (further discussion will be presented
in the last section). This analysis advocates the finial states $\phi(1020)$$f_0(980)$,
$\phi(1020)$$f_0(500)$, $\phi(1680)$$f_0(500)$,
$\omega(782)$$f_0(980)$, $\omega(782)$$f_0(500)$,
$\omega(1420)$$f_0(500)$, $\omega(1650)$$f_0(500)$,
$h_1(1170)$$\eta$, $h_1(1170)$$\eta'$, $\omega(782)$$f_1(1285)$
which satisfy all the constraints from matching concerned quantum numbers.

3. In the simple quark model  $(s\bar s)_{1^{--}}$ can be
decomposed into $c_1\phi(1020)+c_2\omega(782)$ where the values of
$c_1\simeq 1$ and $c_2\simeq 0$ are estimated by fitting the decay
rates of $\phi(1020)\to K^+ K^-$ and $\phi(1020)\to
\pi^+\pi^-$\cite{PDG2016}. In this picture, $\omega$ only contains
a very tiny fraction of strange flavor, thus those modes involving
$\omega(782)$ in the aforementioned channels would have a very small
probability to occur via the fall-apart mechanism directly
but the channel $\phi(2170)\to f_0(980)\omega$ still has a chance
to be measured, which we will discuss latter.

If $f_0(980)$ and $f_0(500)$ are two normal
mesons\cite{Scadron:1982eg,Klabucar:2001gr,Cheng:2002ai,Ke:2009ed,Ke:2017wni},
the decomposition follows $(s\bar
s)_{0^{++}}=c_1'f_0(980)+c_2'f_0(500)$. Moreover, another relation
is $(s\bar s)_{0^{-+}}=c_1''\eta+c_2''\eta'$.  For the $1^{+-}$
quantum system, the only candidate is $(s\bar
s)_{1^{+-}}=h_1(1170)$. With those decompositions we may estimate
the corresponding decay rates of $\phi(2170)$ into the final
products involving those mesons via the fall-apart mechanism.

It is widely accepted that if the fall-apart mechanism exists,
the dominant decay processes should be determined
via this mechanism. Thus we can estimate the decay rates of $\phi(2170)$
roughly by inputting the coefficients
of relevant decompositions and the relations are listed in the following table.

\begin{table*}
\caption{Some factors for the decay $\phi(2170)\to $two
mesons}\label{Tab:t11}
\begin{ruledtabular}
\begin{tabular}{cccccccccc}
  decay mode    &  $\phi$$f_0(980)$ & $\phi$$f_0(500)$  & $h_1$$\eta$&$h_1$$\eta'$  & $\phi$$\eta$&$\phi$$\eta'$& $\omega$$f_0(980)$& $K_1(1270)$$K$& $K_1(1400)$$K$\\\hline
  color factor  & $\frac{2}{\sqrt{3}}$   &  $\frac{2}{\sqrt{3}}$    &   $\frac{2}{\sqrt{3}}$&  $\frac{2}{\sqrt{3}}$&
    $\frac{2}{\sqrt{3}}$&  $\frac{2}{\sqrt{3}}$&  $\frac{2}{\sqrt{3}}$  &  $\frac{4}{3\sqrt{3}}$&  $\frac{4}{3\sqrt{3}}$\\
   spin factor  & $2\sqrt{2}$   &  $2\sqrt{2}$     &  $2\sqrt{2}$ & $2\sqrt{2}$   &  $2\sqrt{2}$ & $2\sqrt{2}$& $2\sqrt{2}$& $4$& $4$\\
flavor factor  & $c_1'$    &  $c_2'$    & $c_1''$ & $c_2''$ & $c_1''$ & $c_2''$& $c_1'$ &1 &1\\
phase space factor\footnote{The partial decay width is
$d\Gamma=\frac{1}{32\pi^2}\frac{|\mathbf{p}|}{m^2}|\mathcal{M}|^2d\Omega$,
where $\mathcal{M}$ is the hadronic transition amplitude.
Supposing it is irrelevant to the solid angle, one can easily
integrate the width over the phase space factor
$\frac{1}{8\pi}\frac{|\mathbf{p}|}{m^2}$. } & 0.0036 & 0.0056 &
0.0054 &0.0018 &0.0062 &0.0038&0.0053&0.0050&0.0041
\end{tabular}
\end{ruledtabular}
\end{table*}

Relevant factors for the decays of $\phi(2170)$ are listed in Tab.
\ref{Tab:t11}. There exists an unknown
factor $g_{FA}$ which is the parameter corresponding to the fall-apart mechanism,
and it should be universal for all the processes.

At present, accurate values of the coefficients $c_1'$,  $c_2'$, $c_1''$ and
$c_2''$ cannot be obtained from data, because so far there
are no measurements with sufficient precision on the relevant processes available
yet. However, we can make rough estimates using the information
we have so  far. That is what we are doing below.

4. $\phi\to
f_0(980)\gamma$ exists, but $\phi\to f_0(500)\gamma$ does not \cite{PDG2016}, the
fact implies $c_1'\simeq 1$. Of course, a possibility is that $f_0(500)$ is a rather wide
resonance, such a radiative decay would be hard to observe. Anyhow, one can roughly
assert that
the channel $\phi(2170)\to \phi f_0(500)$ may be of small probability
to be found and then we set $c_1'= 1$.

Both $\Gamma(D_s\to \eta'\pi^+)$ and $\Gamma(D_s\to \eta\pi^+)$
have been measured, and one can obtain the ratio of the rates of
the two channel as $\Gamma(D_s\to \eta'\pi^+)/\Gamma(D_s\to
\eta\pi^+)=2.32$. Taking account the phase space difference 0.82,
we obtain the ratio $c_2''/c_1''=1.68$. Then we can use the ratio
and the required normalization condition $|c_1''|^2+|c_2''|^2=1$
to determines the modules of coefficients $c_1''$ and $c_2''$. The
advantage of using the ratio instead of the widths enables us to
avoid some experimental errors. Finally  $|c_2'' |= 0.89$ and
$|c_1''| = 0.51$ are achieved. However, in this scenario, we
cannot determine the relative phase between $c_1''$ and $c_2''$.
Using these values, an estimate on the ratios is made as:
$\Gamma(\phi(2170)\to\phi f_0(980))$ : $\Gamma(\phi(2170)\to
h_1\eta)$:$\Gamma(\phi(2170)\to h_1\eta')\simeq$1:0.4:0.4. We
suggest to experimentally search the two channels $\phi(2170)\to
h_1\eta$ and $\phi(2170)\to h_1\eta$ because they do have
substantial branching ratios and should be ``seen" according to
our prediction.

5. Even though $\omega f_0(980)$ cannot be produced via the
fall-apart mechanism the $s$  from the diquark and and $\bar s$
from the anti-diquark of the tetraquark $ss\bar s\bar s$ can
annihilate into $u\bar u$ or $d\bar d$  while the rest $s\bar
s$-pair forms $f_0(980)$ (see Fig.\ref{t1}),  by this picture,
$\phi(2170)\to\omega f_0(980)$ still can be seen in the
experiment, but comparing with $\psi(2170) \to \phi f_0(980)$ the
mode should be suppressed  by $\alpha_s^2$ and an additional color
matching factor.

It is noted that transition  $D_s \to \omega\pi^+$
occurs via weak interaction, namely the charm-quark turns into $s+u\bar d$,
and the spectator $\bar s$ joins the produced $s-$quark, thus the
$s\bar s$-pair
annihilates into $u\bar u$ or $d\bar d$. Due to the similarity, phenomenalogically,
we can use the data of
$D_s \to \phi\pi^+$ and $D_s \to \omega\pi^+$ to predict the width
of $\phi(2170)\to\omega f_0(980)$. Using the ratio $\Gamma(D_s \to
\omega\pi^+)/\Gamma(D_s \to \phi\pi^+)=0.053$ and taking the
corresponding phase factors into account, we have
$\Gamma(\phi(2170)\to\omega f_0(980)):\Gamma(\phi(2170)\to\phi
f_0(980))\approx0.068:1$.

\begin{figure}
\begin{center}
\scalebox{0.4}{\includegraphics{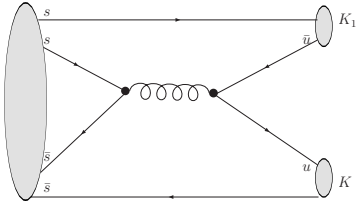}}
\end{center}
\caption{The Feynman diagram for $ss\bar s \bar s\to K_1\bar K$
transitions}\label{t2}
\end{figure}

6. Along the same line, since there are no valence $u\bar u$ or $d\bar d$
components in the tetraquark
$ss\bar s\bar s$, $\psi(2170)$ cannot fall apart into $K_1(1270)K$
or $K_1(1400)K$. In order to produce $K_1(1270)K$ or
$K_1(1400)K$ an $s\bar s$-pair in tetraquark annihilates into
$u\bar u$ or $d\bar d$. The leading Feynman diagram is Fig.
\ref{t2}.
The color and spin factors are presented in Tab. \ref{Tab:t11}
and the production process is somehow similar to $\phi(2170)\to\phi
f_0(980)$, but is suppressed by $\alpha_s^2$.
For $c\bar c$ system $\alpha_s$
is about $0.39$\cite{Eichten:1978tg}, whereas for the $s\bar s$ case $\alpha_s$
may be slightly larger, but the suppression exists. Moreover,
there are twice-color matching (at initial and final sides), thus an extra
factor $g_{FA}$ is introduced.


7. If we
set $\alpha_s\sim0.5$ and $g_{FA}\sim 1$ we expect
$\Gamma(\phi(2170)\to K_1(1270)K) :\Gamma(\phi(2170)\to\phi
f_0(980))\sim0.31:1$. In terms of
$\Gamma(\Upsilon(4S)\to\Upsilon(1S)\pi\pi)\approx\Gamma(\Upsilon(4S)\to\Upsilon(2S)\pi\pi)
$  we estimate $\Gamma(\phi(2170)\to K_1(1270)K)$ to be close to
$\Gamma(\phi(2170)\to K_1(1400)K)$. We would ask whether
 $ K^*(890)^0\bar K^*(890)^0$, $K^+K^-$ and
$K^0_LK^0_S$ can be experimentally measured? Since the relative
orbital angular momentum between the daughter mesons is $l=1$,
then since the reactions occur near the threshold, the 3-momentum
is small, thus the $p$-wave suppression would remarkably reduce
the production rate, comparing to $s$-wave case. A rough estimate
of the suppression factor is $\frac{\mathbf{p}^2}{m^2}\sim 0.08$.
Moreover, to take into account additional factors which may
affect evaluation, we adopt the suppression factor for the $p$-wave
using the data $\Gamma(\psi(2S)\to \eta_c\gamma)$ and
$\Gamma(\psi(2S)\to \chi_0\gamma)$  and where the three-momentum
of finial mesons is close to that in $\phi(2170)\to K^*(890)^0\bar
K^*(890)^0$. With the ratio $\Gamma(\psi(2S)\to
\eta_c\gamma)/\Gamma(\psi(2S)\to \chi_0\gamma)$ the production rate of
$p$-wave in the case of $\psi(2170)$ is suppressed and is about $
0.034$ times smaller than that for $s$-wave. Thus,
$\Gamma(\phi(2170)\to K^*(890)^0\bar
K^*(890)^0):\Gamma(\phi(2170)\to\phi f_0(980))$ is estimated as
$\sim0.01:1$. Meanwhile $\Gamma(\phi(2170)\to K^+K^-)$ and
$\Gamma(\phi(2170)\to K^0_LK^0_S)$ should be close to
$\Gamma(\phi(2170)\to K^*(890)^0\bar K^*(890)^0)$.

8. For other $p$-wave decays of $\phi(2170)$ into $\phi\eta$ and
$\phi\eta'$ incorporating the phase factors we estimate
$\Gamma(\phi(2170)\to\phi\eta):\Gamma(\phi(2170)\to\phi\eta'):\Gamma(\phi(2170)\to\phi
f_0(980))\sim0.015:0.025:1$.

9. $f_0(980)$ may also considered as a molecular
state\cite{Weinstein:1982gc} or a tetraquark\cite{Jaffe:1976ig},
if so, the picture would be slightly different and the relevant
Feynman diagram is shown in the following figure \ref{t3}.
\begin{figure}
\begin{center}
\scalebox{0.4}{\includegraphics{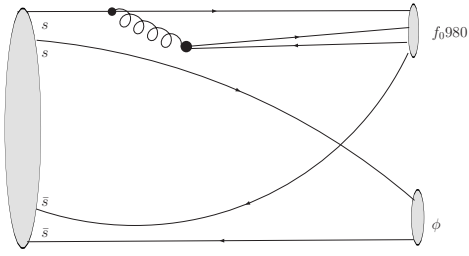}}
\end{center}
\caption{The Feynman diagram for $ss\bar s \bar s\to \phi
f_0(980)$ transition as $f_0(980)$ is regarded as a multi-quark
state}\label{t3}
\end{figure}
In this case $\phi(2170)\to \phi f_0(980)$ is suppressed by
$\alpha_s^2$ comparing to the aforementioned case where $f_0(980)$
is supposed to be a normal meson.  Now the ratio
$\Gamma(\phi(2170)\to\phi f_0(980))$:$\Gamma(\phi(2170)\to
h_1\eta)$:$\Gamma(\phi(2170)\to h_1\eta')$ would be close to
1:1.6:1.6. However, as well understood, $f_0(980)$ may be a
mixture of $s\bar s$ state and a multi-quark state, thus according
to our estimate, one can roughly evaluate the fraction of each
constituent, and it would answer a long standing question about
the identity of $f_0(980)$. Obviously precise measurement on
$\phi(2170)\to \phi f_0(980)$ would be very helpful.


10. At last we can estimate the results if $\phi(2170)$ is
an excited $q\bar q s\bar s$ tetraquark ($q=u, d$) as suggested.
Naturally, the following decay portals would dominate the total width of $\phi(2170)$, they are:
$q\bar q$+$s\bar s$ and $q\bar
s$+$s\bar q$ which can be realized via the fall apart mechanism. In this case we
expect $\phi(2170)\to\phi f_0(500)$ to be main decay channel
rather than $\phi(2170)\to\phi f_0(980)$. $\Gamma(\phi(2170)\to
K_1(1270)\bar K)$ and $\Gamma(\phi(2170)\to K_1(1400)\bar K)$
should be close to $\Gamma(\phi(2170)\to\phi f_0(500))$. In this
case $\phi(2170)\to K^*(890)^0\bar K^*(890)^0$, $\phi(2170)\to
K^+K^-$ and $\phi(2170)\to K^0_LK^0_S$ only receive a
$p$-wave suppression, but not color rearrangement suppressions,
different from the aforementioned case.
\section{Discussions and Conclusion}

With the study on the multi-quark structures stepping deeper and
deeper, many unanswered puzzles in this stimulating field have
emerged, namely sharp contradiction between theoretical prediction
and experimental observation reminds us that our understanding of
the exotic hadrons is far away from  satisfaction. For example,
many theoretical models confirm existence of $X(5568)$, however,
all experimental collaborations offered negative reports
\cite{Aaij:2016iev,Sirunyan:2017ofq,Aaltonen:2017voc,Aaboud:2018hgx}
except the D0 collaboration \cite{D0:2016mwd}. To compromise the
contradiction between theory and experiment, we suggested that a
destructive interference between the molecular state and
tetraquark suppressed the concerned decay
portals\cite{Ke:2018stp}. Indeed, it is a bold conjecture and
needs further verification by both of theoretical calculations and
more accurate experimental measurements.

The
decay modes of $\phi(2170)$ imply that the assignment of being an
excited state of $\phi$ is disfavored. Some authors suggested that
it should be an exotic state. More concretely, its mass and decay
behaviors hint that it may be an $ss\bar s\bar s$ tetraquark
state. Such a structure is special
because it may decay via the so-called fall-apart
mechanism into hadrons which possess dominantly strange
constituents. Employing the fall apart mechanism we estimate the
decay modes of $\phi(2170)$ which are supposed to be its dominant
portals. If $f_0(980)$  is a simple meson with $s\bar s$
structure, our estimate show that $\Gamma(\phi(2170)\to\phi
f_0(980))$:$\Gamma(\phi(2170)\to h_1\eta)$:$\Gamma(\phi(2170)\to
h_1\eta')\simeq$1:0.4:0.4, $\Gamma(\phi(2170)\to\omega
f_0(980)):\Gamma(\phi(2170)\to\phi f_0(980))\approx0.068:1$ and
$\Gamma(\phi(2170)\to K_1(1270)K) :\Gamma(\phi(2170)\to
K_1(1400)K):\Gamma(\phi(2170)\to\phi f_0(980))\sim0.31:0.31:1$. In
this case $\Gamma(\phi(2170)\to K^*(890)^0\bar K^*(890)^0)$,
$\Gamma(\phi(2170)\to K^+K^-)$ and $\Gamma(\phi(2170)\to
K^0_LK^0_S)$ are suppressed by about two orders comparing with
$\Gamma(\phi(2170)\to \phi f_0(980))$.

If
$f_0(980)$ is a four-quark state, the decay
$\phi(2170)\to\phi f_0(980)$ is suppressed and the ratio
$\Gamma(\phi(2170)\to\phi f_0(980))$:$\Gamma(\phi(2170)\to
h_1\eta)$:$\Gamma(\phi(2170)\to h_1\eta')\simeq$ would be close to
1:1.6:1.6.

Supposing $\phi(2170)$ is an excited $q\bar q s\bar s$ tetraquark
($q=u, d$) $\phi(2170)\to\phi f_0(500)$, $\phi(2170)\to
K_1(1270)\bar K$ and $\phi(2170)\to K_1(1400)\bar K$ is
expected to be the main decay channels. Even though $\phi(2170)\to
K^*(890)^0\bar K^*(890)^0$, $\phi(2170)\to K^+K^-$ and
$\phi(2170)\to K^0_LK^0_S$ are $p$-wave suppressed modes,
since the three-momenta for these channels are not too
small, they
should be observed experimentally.


Along with all  other subjects in the hadron physics, a better
understanding of the exotic state structure and their production
and decay mechanisms are badly needed. We all know that the fundamental theory
of strong interaction is QCD, however, the non-perturbative QCD which governs
the hadron physics is still not understood yet, so that various reasonable
phenomenological models are adopted by researchers. The study on exotic state
may help us to gain more information about quark model and non-perturbative QCD.
As discussed in the text, $\phi(2170)$ is a special case worth of concern.

We suggest to measure all
decay modes of $\phi(2170)$ because the data will inform us of its
assignment. If the data can decide it to be an $ss\bar s\bar s$ tetraquark,
just as we mentioned in the introduction, existence of multi-quark states
with only $s$-flavor which is not very heavy is confirmed, and our scope
would be widened.

From our discussion, one can notice that to gain more  solid
knowledge on the structure of exotic states and concerned dynamics
is not easy because many inputs adopted in the computations
possess large errors. It means that accurate data are the
precondition for drawing definite conclusions. So far, the
available facilities cannot offer data with satisfactory accuracy
in the energy range of charm, thus we lay hope on the future
charm-tau factory which is planned to be built in China. Since the
luminosity of the new facility would be enhanced by several orders
than that of BEPC II, and some new detection technique will be
used,  we can be optimistic  that the quality of data will be much
improved and the statistics can reach a very high level. Then, we
may renew our computations based on the more accurate data and
draw definite conclusion not only about $\psi(2170)$ but also many
four-quark states and pentaquarks.

\section*{Acknowledgment}

One of us (Ke) would like to acknowledge a fruitful discussion
with Dr. Wen-Biao Yan. This study was supported by the National
Natural Science Foundation of China (NNSFC) under contract No.
11375128 and 11675082.

\end{document}